\documentclass[compsoc,conference,10pt,times]{IEEEtran}

\usepackage{pifont}
\usepackage{tikz}
\usepackage{listings}
\usetikzlibrary{patterns}
\usetikzlibrary{tikzmark}
\usetikzlibrary{calc}
\usetikzlibrary{fit}
\usetikzlibrary{arrows.meta, positioning}

\usepackage{soul}
\usepackage{inconsolata}

\usepackage{oplotsymbl}
\usepackage[hyphens]{url}
\usepackage[colorlinks,allcolors=black]{hyperref}

\usepackage{pgfplots}
\usepackage{pgfplotstable}
\usepackage{float}

\usepackage{amsmath}
\usepackage{amssymb}
\usepackage{newtxtext} 
\usepackage{ntheorem}
\usepackage{tikz}
\usepackage[lined]{algorithm2e}

\SetDataSty{myargfont}
\SetAlCapSkip{1em} 

\usepackage{graphicx}
\usepackage[table,xcdraw]{xcolor}
\usepackage{multirow}
\usepackage{tabularx}
\usepackage{plai}

\usepackage{booktabs}

\usepackage{relsize} 
\usepackage{circledsteps} 

\usepackage{enumerate}

\newtheorem{definition}{Definition}

\definecolor{blue}{HTML}{5e81ac}
\definecolor{green}{HTML}{a3be8c}
\definecolor{red}{HTML}{bf616a}
\definecolor{darkblue}{HTML}{3b4252}
\definecolor{yellow}{HTML}{ebcb8b}

\lstdefinestyle{overviewlst}{
  style=plai,
  basicstyle=\ttfamily\tiny,
  keywordstyle=\color{blue}, 
  commentstyle=\color{green}, 
  stringstyle=\color{red}, 
  keywordstyle={[2]\color{red}},
  keywordstyle={[3]\color{violet}}, 
  backgroundcolor=\color{green!20},
  aboveskip={0pt},
  belowskip={0pt},
}

\lstdefinestyle{greenblock}{
  style=overviewlst,
  backgroundcolor=\color{green!20}
}

\lstdefinestyle{redblock}{
  style=overviewlst,
  backgroundcolor=\color{red!20}
}

\lstdefinestyle{grayblock}{
  style=overviewlst,
  backgroundcolor=\color{gray!10}
}

\lstdefinestyle{assemblyblock}{
  style=plai,
  basicstyle=\ttfamily\footnotesize,
  keywordstyle=\color{blue}, 
  commentstyle=\color{green}, 
  stringstyle=\color{red}, 
  keywordstyle={[2]\color{red}},
  keywordstyle={[3]\color{violet}}, 
  backgroundcolor=\color{gray!5}
}

\usepackage{cite}
\usepackage{amsmath,amssymb,amsfonts}
\usepackage{algorithmic}
\usepackage{graphicx}
\usepackage{textcomp}
\usepackage{bmpsize}
\usepackage{xcolor}
\usepackage{lipsum}

\def\BibTeX{{\rm B\kern-.05em{\sc i\kern-.025em b}\kern-.08em
    T\kern-.1667em\lower.7ex\hbox{E}\kern-.125emX}}

\newcommand*\circled[1]{{\smaller\Circled{\textrm{#1}}}}

\begin{document}

\title{Match\,\&\,Mend: Minimally Invasive Local Reassembly for Patching N-day Vulnerabilities in ARM Binaries}

\author{\IEEEauthorblockN{Sebastian Jänich}
\IEEEauthorblockA{\textit{LMU Munich}\\
Munich, Germany \\
sebastian.jaenich@lmu.de}
\and
\IEEEauthorblockN{Merlin Sievers}
\IEEEauthorblockA{\textit{LMU Munich}\\
Munich, Germany \\
merlin.sievers@lmu.de}
\and
\IEEEauthorblockN{Johannes Kinder}
\IEEEauthorblockA{\textit{LMU Munich}\\
Munich, Germany \\
johannes.kinder@lmu.de}
}

\maketitle

\begin{abstract}
    Low-cost Internet of Things (IoT) devices are increasingly popular but often insecure due to poor update regimes. As a result, many devices run outdated and known-vulnerable versions of open-source software. 
    We address this problem by proposing to patch IoT firmware at the binary level, without requiring vendor support.
    In particular, we introduce minimally invasive local reassembly, a new technique for automatically patching known (n-day) vulnerabilities in IoT firmware. Our approach is designed to minimize side effects and reduce the risk of introducing breaking changes. We systematically evaluate our approach both on 14 vulnerabilities within the controlled environment of the MAGMA benchmarks, as well as on 30 real-world Linux-based IoT firmware images from the KARONTE dataset. Our prototype successfully patches 76\% of targeted vulnerabilities in MAGMA and 96\% in the firmware dataset.
\end{abstract}


\section{Introduction}

Internet of Things (IoT) devices, from WiFi routers over smart cameras to robot vacuums, are widely deployed in homes and businesses. Their firmware usually combines open-source components with some proprietary code implementing device- or application-specific functionality. A popular and flexible architecture design choice is Linux for 32-bit ARMv7, which can form a basis for minimal to complex embedded systems distributions~\cite{yocto-pr}.

Especially for low-cost devices, security is often not a priority, so they are insufficiently hardened and receive updates only irregularly or never at all. This leads to IoT devices running outdated software containing known vulnerabilities and potentially becoming targets or entry points for attackers~\cite{NinoL0L0G24}. 
This issue is wide-spread; by scanning 16 million home networks, Kumar et al.~\cite{KumarSCGAK19} found 40\% of households worldwide to have at least one vulnerable smart device. 
While recent policy initiatives (e.g., in the European Union~\cite{cra2024}) require vendors to provide updates, direct sales of low-cost devices to end users circumvent even existing safety standards, so it is highly likely that such devices will continue threatening the security of the networks they are deployed on.

In this paper, we focus on a purely technical solution to vulnerable IoT devices, based on the idea of simply patching any known (n-day) vulnerabilities \textit{without support from the manufacturer}.
Key to our approach is a highly localized static binary reassembly algorithm that patches just the vulnerability, while ensuring consistency and leaving the rest of the firmware untouched. 
Our guiding principle in designing our solution is to avoid introducing any breaking changes to the firmware.

Clearly, an initial obstacle is to extract the firmware from the device. Possible access paths depend on the specific device, but they follow common patterns~\cite{VasileOC18}. The fully patched firmware then has to be reinstalled on the device in a similar fashion. While some devices do use signature schemes that prevent installing custom firmware (e.g., Amazon Echo, Apple Homepod), many update mechanisms are not secured~\cite{IbrahimCB23,WuWWZLH00024}. 
Additionally, physical access opens many more possibilities, e.g., through exposed debug interfaces such as UART and JTAG~\cite{SuR22,dark}, or via backup mechanisms like flashing from an SD card~\cite{allwinner}.
In particular, we argue that on low-cost devices the same type of lax security posture that makes our approach to patching \emph{necessary} also makes it \emph{feasible}.

The ideal solution would be to obtain the original source code and build environment for the device, update any included open-source components to fix known vulnerabilities, rebuild the firmware and upload it to the device. However, without the cooperation of the manufacturer, we lack critical components for building the firmware. While the code of the open-source components is available, we will miss any custom closed-source applications added by the manufacturer; any custom patches that were applied to a standard release of the open-source components; and the specific build configuration including compile-time switches.
For instance, OpenSSL has over 20 compile-time options that enable or disable various features and protocols. If we were to replace the entire OpenSSL package because of a vulnerability, we would have to reconstruct the exact same combination of options to reduce any differences in behavior, including resource consumption; even for individual functions, this is a challenging problem~\cite{DuanBJAXISL19}.

Therefore, we see \emph{micro-patching} of known vulnerabilities at the binary level as a promising path to improve the security of low-cost devices.
In this paper, we focus on the problem of flexibly rewriting and reassembling the patch in a target binary; a complete end-to-end solution additionally requires applying existing techniques for identifying open-source libraries in firmware~\cite{libdb} and finding known vulnerabilities~\cite{DuanBXKL17,DavidPY18}. In our prototype, we employ Intel's publicly available CVE-bin tool~\cite{tool:CVE-bin-tool} for this purpose.

Binary rewriting is a challenging problem in itself. Existing approaches involve either static~\cite{KimKCKSZX17,DuckGR20,egalito,BaumanLH18} or dynamic~\cite{DuanBJAXISL19,He22,NieslerSD21,SalehiP24} techniques to modify the binary code. A survey by Wenzl et al.~\cite{WenzlMUW19} finds that work so far predominantly focuses on the x86 architecture, leaving 32-bit ARM binaries, typical for IoT devices, with fewer solutions.
To the best of our knowledge, in this paper we are the first to propose a minimally invasive local reassembly approach for an automated patching process.
Crucially, we require a persistent solution where the code is written into the binary itself. Unlike dynamic approaches such as OSSPatcher, which pioneered runtime patching on Android~\cite{DuanBJAXISL19}, we cannot rely on intercepting the loading or linking process on general IoT devices, where the global flow of control may be unknown. 
Our problem is also harder than other common static binary rewriting tasks: the code being rewritten can interact heavily with the rest of the binary, including references to functions and global variables.

However, our setting offers the unique advantage that we do have \emph{some} source code for the binary we are rewriting; we just may not have the exact same source code, and the compilation of the target is outside our control. 
We leverage our knowledge about the code to identify functions to be patched and  match basic blocks such that we can reconstruct code and data references.

Match\&Mend combines static binary analysis and local reassembly of ARM32 binary code to produce automated, minimally invasive micropatches.
This yields a practical local reassembly pipeline, transplanting instructions from one binary into another with recomputed relative offsets and relocations avoiding global relinking or full binary recompilation, thereby reducing layout perturbation and side effect risk.
In summary, the main contributions are: 
\begin{itemize} 
  \item A novel concept for local reassembly by resolving a scoped symbolic mapping of instruction‑level references (e.g., PC‑relative loads, literal pools, GOT/PLT entries, and function branch targets) to concrete addresses/offsets in the host binary~(\autoref{sec:definitions}).
  \item An algorithm for minimally invasive local reassembly of micro-patches in ARM binaries, to effectively patch known vulnerabilities in IoT firmware~(\autoref{sec:rewriting}).
  \item The implementation of this algorithm in \emph{Match \& Mend}, a working prototype system for automated binary patching in ELF binaries~(\autoref{sec:implementation}). 
\end{itemize}
We thoroughly evaluate our prototype on both the MAGMA benchmark and a dataset of real-world firmware images, demonstrating that Match \& Mend can successfully patch firmware vulnerabilities in practice~(\autoref{sec:evaluation}).
All source code and data are available on GitHub.\footnote{\url{https://github.com/lmu-plai/match-and-mend}}

\section{Problem Definition}
\label{sec:scope_challenges}
We begin by outlining the scope of the automated binary patching problem~(\autoref{sec:scope}) and identifying the core challenges~(\autoref{sec:challenges}).

\subsection{Scope}
\label{sec:scope}

Binary patching is the process of fixing a vulnerability in a binary executable without changing its intended functionality.
In the following, we informally define our interpretations of relevant terms to help explain our approach.
We consider a \emph{vulnerability} in a binary to be the set of instructions and data that corresponds to the minimal set of statements in source code that enables the attacker to potentially exploit the binary, e.g., an unchecked access of an array in C. For the same source code, there can be many instances of the vulnerability in binary form. Similarly, a \emph{patch} in binary form is the set of instructions and data that correspond to the minimal set of statements in source code that need to be modified or added to eliminate the vulnerability without changing the originally intended behavior, e.g., adding a check before accessing an array in C.

The scope of our binary patching problem is defined as follows: there is a Linux-based IoT firmware that contains open-source libraries. Among these libraries is one with a known and documented n-day vulnerability. Furthermore, a patched version of the library's source code is available. The task is to determine how to patch the vulnerable binary in an automated and minimally invasive way.

\subsection{Challenges}
\label{sec:challenges}

The main goal of existing rewriting techniques is to enable instrumentation of binaries for various purposes such as fuzzing, hardening, or analysis. The core challenges that need to be solved to correctly rewrite a binary, such as
pointer detection, telling code from data and handling indirect control flow transfers, all without introducing breaking changes or too much overhead, can be solved using existing techniques. We identified new challenges specific to the problem of binary patching:

\smallskip\noindent
\textbf{(C1)\enspace Patch Identification}:
To be able to correctly patch a vulnerability, a system needs to identify a patch in binary code that removes the vulnerability. 
\smallskip

\smallskip\noindent
\textbf{(C2)\enspace Vulnerability Identification}:
The binary code containing the vulnerability has to be identified. Even when the vulnerability is known in source code form, the exact source code and flags for compilation generally are not.

\smallskip\noindent
\textbf{(C3)\enspace Control Flow Integration}: The control flow of the patch has to be integrated into the control flow of the vulnerable code. This involves jump and call redirection to connect the patch with the target binary.
  
\smallskip\noindent
\textbf{(C4)\enspace Data Flow Integration}: Also the data flow of the patch needs to be integrated into the data flow of the vulnerable code. This may include adding new data to the binary.
  
\smallskip\noindent
\textbf{(C5)\enspace Encoding Modified Instructions}: During reassembly, instructions and their sizes (depending on encoding format) might change. However, due to relative addressing, the relative distances between instructions need to be preserved to not introduce breaking changes.

\begin{figure*}[!t]
  \centering
  \begin{tikzpicture}[>=stealth,font=\sffamily\notsotiny,
    cfgnode/.style={shape=rectangle, draw, color=black!80, align=center, fill=gray!20, minimum height=0.25cm, inner sep=0pt, minimum width=9.55pt, text width=0.01cm},
    cfgnode_short/.style={shape=rectangle, draw, color=black!80, align=center, fill=gray!20, minimum height=0.14cm, inner sep=0pt, minimum width=9.55pt, text width=0.01cm},
    cfgnode_line/.style={shape=rectangle, draw, color=black!80, align=center, fill=gray!20, inner sep=0pt, minimum height=0.3mm, minimum width=9.55pt, text width=0.01cm},
    cfgblock/.style={rectangle,draw,text centered,inner sep=2pt,color=black!80, text=white, anchor=north},
    titlebar/.style={rectangle,text centered,inner sep=0pt,minimum height=4mm,color=black!80, text=black, fill=yellow!20},
    subtitlebar/.style={rectangle,draw,text centered,inner sep=0pt,minimum height=4mm,color=black!80, text=black, anchor=north},
    ]
          
    \node[titlebar,minimum width=2.5cm] (vulnerable_binary) at (8, 5){Vulnerable Library};
    \node[subtitlebar,minimum width=2.5cm] (vulnerable_bottom) at (vulnerable_binary.south) {\lstinline+libpng 1.0.65+};
    \node[cfgblock,minimum width=2.5cm,minimum height=76pt] (vulnerable_binary_bottom) at (vulnerable_binary.north){};
    
    \node[titlebar,minimum width=2.5cm] (patch_binary) at (8, 9){Patch Library};
    \node[cfgblock,minimum width=2.5cm,minimum height=76pt] (patch_binary_bottom) at (patch_binary.north){};
    \node[subtitlebar,minimum width=2.5cm]  at (patch_binary.south) {\lstinline+libpng 1.0.66+};

    \node[titlebar,minimum width=2.8cm] (patched_firmware) at ($(patch_binary.north east)+(23mm,-20mm)$){Patched Vulnerable Library};
    \node[cfgblock,minimum width=2.8cm,minimum height=87pt] (patched_firmware_bottom) at (patched_firmware.north){s};
    \node[subtitlebar,minimum width=2.8cm]  at (patched_firmware.south) {\lstinline+libpng 1.0.65_patch+};


    \begin{scope}[>=stealth, node distance=0.5cm,
      shift={(vulnerable_binary.south)}, yshift=-18pt]
      \node[cfgnode] (start) [] {};
      \node[cfgnode, fill=red] (checkstart) [below right of=start] {};
      \node[cfgnode] (checkleft) [below left of=checkstart, xshift=0.08cm] {};
      \node[cfgnode] (left) [left of=checkleft] {};
      \node[cfgnode, fill=red] (checkright) [below right of=checkstart, xshift=-0.08cm] {};
      \node[cfgnode](checkend) [below left of=checkright, xshift=0.08cm] {};
      \node[cfgnode] (end) [below left of=checkend] {};
      
      \draw[color=gray] (start.south) -- (left.north);
      \draw[color=gray] (start.south) -- (checkstart.north);
      \draw[color=gray] (left.south) -- (end.north);
      \draw[color=gray] (checkstart.south) -- (checkleft.north);
      \draw[color=red] (checkstart.south) -- (checkright.north);
      \draw[color=gray] (checkleft.south) -- (checkend.north);
      \draw[color=gray] (checkright.south) -- (checkend.north);
      \draw[color=gray] (checkend.south) -- (end.north);

    \end{scope}
    \node[left of=left,solid, xshift=12pt]{\circled{1}};
  
    \node[anchor=north east] (vuln_check) at ($(vulnerable_binary.north west)-(10mm,-3mm)$) {%
    \begin{minipage}{4cm}%
    \begin{lstlisting}[style=redblock]
  1024: mov   r5, r2
  1026: cmp   r5, #0
  1028: bne   #0x1034
    \end{lstlisting}%
    \end{minipage}%
    };
  
    \node[anchor=north] (vuln_else) at (vuln_check.south){%
    \begin{minipage}{4cm}%
    \begin{lstlisting}[style=redblock]
  102A: ldr   r0, [pc, #0x38]
  102C: add   r0, pc
  102E: bl    png_error
  1032: b     #0x1036
    \end{lstlisting}%
    \end{minipage}%
    };

    \node[anchor=north] (vuln_then) at (vuln_else.south){%
    \begin{minipage}{4cm}%
    \begin{lstlisting}[style=grayblock]
  1034: adds  r5, #1
    \end{lstlisting}%
    \end{minipage}%
    };

    \node[anchor=north] (vuln_tail) at (vuln_then.south){%
    \begin{minipage}{4cm}%
    \begin{lstlisting}[style=grayblock]
  1036: ldr   r3, [pc, #0xA2]
  103A: str   r6, [r3, r5]
    \end{lstlisting}%
    \end{minipage}%
    };

    \node (zoom_vuln) [shape=rectangle,draw,dashed,color=gray, minimum width=0.5cm, minimum height=10.5mm, text width=0.80cm,yshift=3.75pt, xshift=-7.5pt] at (checkright.south) {};
	\node(zoom-big_vuln)[draw,dashed,color=gray,fit=(vuln_check)(vuln_tail)] at ($(vuln_check.north)!.5!(vuln_tail.south)+(0pt,1.8pt)$) {};
    
    \draw[gray, dashed] (zoom_vuln.north west) -- (zoom-big_vuln.north east);
    \draw[gray, dashed] (zoom_vuln.south west) -- (zoom-big_vuln.south east);

  
    \begin{scope}[>=stealth, node distance=0.5cm, shift={(patch_binary.south)}, yshift=-18pt ]
      \node[cfgnode] (start) [] {};
      \node[cfgnode, fill=green] (patchstart) [below right of=start] {};
      \node[cfgnode] (patchleft) [below left of=patchstart, xshift=0.08cm] {};
      \node[cfgnode] (left) [left of=patchleft] {};
      \node[cfgnode, fill=green] (patchright) [below right of=patchstart, xshift=-0.08cm] {};
      \node[cfgnode](patchend) [below left of=patchright, xshift=0.08cm] {};
      \node[cfgnode] (end) [below left of=patchend] {};
      
      \draw[color=gray] (start.south) -- (left.north);
      \draw[color=gray] (start.south) -- (patchstart.north);
      \draw[color=gray] (left.south) -- (end.north);
      \draw[color=gray] (patchstart.south) -- (patchleft.north);
      \draw[color=green] (patchstart.south) -- (patchright.north);
      \draw[color=gray] (patchleft.south) -- (patchend.north);
      \draw[color=gray] (patchright.south) -- (patchend.north);
      \draw[color=gray] (patchend.south) -- (end.north);
    \end{scope}
    \node[left of=left,solid, xshift=12pt]{\circled{1}};

    \node[anchor=north] (patch_check) at ($(vuln_check)+(0pt,43mm)$) {%

    \begin{minipage}{4cm}%
    \begin{lstlisting}[style=greenblock]
  4014: mov   r5, r2
  4016: cmp   r5, #0
  4018: bgt   #0x4026
    \end{lstlisting}
    \end{minipage}
    };
     
    \node (patch_else) [anchor=north]at (patch_check.south){%
    \begin{minipage}{4cm}
    \begin{lstlisting}[style=greenblock]
  401A: ldr   r0, [pc, #0x40]
  401C: movs  r5, #1
  401E: add   r0, pc
  4020: blx   png_error
  4024: b     #0x4028
    \end{lstlisting}
    \end{minipage}
    };

    \node (patch_then) [anchor=north]at (patch_else.south){%
    \begin{minipage}{4cm}
    \begin{lstlisting}[style=grayblock]
  4026: adds  r5, #1
    \end{lstlisting}
    \end{minipage}
    };

    \node (patch_tail) [anchor=north]at (patch_then.south){%
    \begin{minipage}{4cm}
    \begin{lstlisting}[style=grayblock]
  4028: ldr   r3, [pc, #0xB0]
  402A: str   r6, [r3, r5]
    \end{lstlisting}
    \end{minipage}
    };

    \node(zoom_patch) [shape=rectangle,draw,dashed,color=gray, minimum width=0.5cm, minimum height=10.5mm, text width=0.80cm,yshift=3.75pt, xshift=-7.5pt] at (patchright.south) {};
	\node(zoom-big_patch)[draw,dashed,color=gray,fit=(patch_check)(patch_tail)] at ($(patch_check.north)!.5!(patch_tail.south)+(0pt,1.8pt)$) {};
  
    \draw[gray, dashed] (zoom_patch.north west) -- (zoom-big_patch.north east);
    \draw[gray, dashed] (zoom_patch.south west) -- (zoom-big_patch.south east);

  
    \begin{scope}[>=stealth, node distance=0.5cm, shift={(patched_firmware.south)}, yshift=-18pt, xshift=-0.5cm ]
      \node[cfgnode] (start) [] {};
      \node[cfgnode] (left) [below left of=start, yshift=-0.5cm] {};
      \node[cfgnode] (end) [below right of=left, yshift=-0.5cm] {};
	  \node[cfgnode] (tail) [above right of=end, xshift=0.25cm] {};
	  \node[cfgnode, fill=red] (dead_else) [above of=tail] {};
	  \node[cfgnode] (then) [left of=dead_else] {};
	  \node[cfgnode, fill=green] (patch_else) [right of=dead_else] {};
	  \node[cfgnode, fill=green] (patch_if) [above of=dead_else, xshift=2mm, yshift=-1mm] {};
	  \node[cfgnode_short, fill=red] (dead_if) [above of=dead_else, xshift=-3mm] {};
	  \node[cfgnode_line, fill=red] (dead_if_line) [above of=dead_if, yshift=-3.8mm] {};

      \draw[color=gray] (start.south) -- (left.north);
      \draw[color=gray] (start.south) -- (dead_if_line.north);
      \draw[color=green] (dead_if_line.east) -- (patch_if.north);
      \draw[color=gray] (left.south) -- (end.north);
      \draw[color=gray] (tail.south) -- (end.north);
      \draw[color=gray] (then.south) -- (tail.north);
      \draw[color=gray] (dead_else.south) -- (tail.north);
      \draw[color=gray] (dead_if.south) -- (then.north);
      \draw[color=red] (dead_if.south) -- (dead_else.north);
      \draw[color=green] (patch_if.south) -- (then.north);
      \draw[color=green] (patch_if.south) -- (patch_else.north);
      \draw[color=green] (patch_if.south) -- (patch_else.north);
      \draw[color=green] (patch_else.south) -- (tail.north);
    \end{scope}
    
    \node[anchor=north west] (final_dead) at ($(patch_check.north -| patched_firmware.east) + (10mm,-10mm) $) {%

    \begin{minipage}{4cm}%
    \begin{lstlisting}[style=redblock]
  1024: @\\hl{bx}@    @\\hl{\#0xF014}@
  1028: bne   #0x1034
    \end{lstlisting}
    \end{minipage}
    };

    \node[anchor=north west] (final_dead_2) at (final_dead.south west){%
    \begin{minipage}{4cm}%
    \begin{lstlisting}[style=redblock]
  102A: ldr   r0, [pc, #0x38]
  102C: add   r0, pc
  102E: bl    png_error
  1032: b     #0x1036
    \end{lstlisting}
    \end{minipage}
    };
      
    \node[anchor=north] (final_patch1) at ($(final_dead_2.south east) - (14mm,0mm)$){%
    \begin{minipage}{4cm}%
    \begin{lstlisting}[style=greenblock]
  F014: mov   r5, r2
  F016: cmp   r5, #0
  F018: bgt   @\\hl{\#0x1034}@
    \end{lstlisting}
    \end{minipage}
    };

    \node[anchor=north] (final_patch2) at (final_patch1.south){%
    \begin{minipage}{4cm}%
    \begin{lstlisting}[style=greenblock]
  F01C: ldr   r0, [pc, @\\hl{\#0x5C}@]
  F01E: movs  r5, #1
  F020: add   r0, pc
  F022: blx   png_error
  F026: b     @\\hl{\#0x1036}@
    \end{lstlisting}
    \end{minipage}
    };
      
    \node[anchor=north] (final_then) at ($(final_patch2.south west) + (14mm,0mm)$){%
    \begin{minipage}{4cm}%
    \begin{lstlisting}[style=grayblock]
  1034: adds  r5, #1
    \end{lstlisting}
    \end{minipage}
    };

    \node[anchor=north] (final_tail) at (final_then.south) {%
    \begin{minipage}{4cm}%
    \begin{lstlisting}[style=grayblock]
  1036: ldr   r3, [pc, #0xA8]
  103A: str   r6, [r3, r5]
    \end{lstlisting}
    \end{minipage}
    };

    \node (zoom_final) [shape=rectangle,draw,dashed,color=gray, minimum width=14.5mm, minimum height=13.6mm,yshift=0.63cm] at (tail.south) {};
    \coordinate (final_shift) at ($(final_tail.south)+(0,3pt)$);
    \node(zoom-big_final)[draw,dashed,color=gray,fit=(final_dead)(final_patch1)(final_shift)] {};

    \draw[gray, dashed] (zoom_final.north east) -- (zoom-big_final.north west);
    \draw[gray, dashed] (zoom_final.south east) -- (zoom-big_final.south west);

    \draw[->] (vulnerable_binary_bottom.east) -- (patched_firmware_bottom.west);
    \draw[->] (patch_binary_bottom.east) -- node[below,solid,yshift=-0.75cm]{\circled{3}}(patched_firmware_bottom.west);

    \draw[<->, ] (vulnerable_binary) -- node[left,solid]{\circled{2}} (patch_binary_bottom)  node[draw, rectangle, minimum width=1cm, color=black, text=black, inner sep=5pt, rounded corners, fill=gray!10, midway, xshift=0.6cm] {Matcher};
  \end{tikzpicture}
  \caption{Schematic overview of the patching process, using the fictional example of a vulnerable \lstinline+libpng 1.0.65+, designed to concisely showcase the challenges and solutions of minimally invasive local reassembly. The vulnerability is patched in \lstinline+libpng 1.0.66+. \circled{1} We identify vulnerable and patched functions in the binary and construct control and data flow graphs; \circled{2} we match basic blocks between the functions using binary diffing and use this information to match references to code and data; \circled{3} we transplant the code from the patched to the vulnerable version, adjusting all code and data references as needed and fixing up shifts from differences in addressing modes.}
  \label{fig:design_overview}
\end{figure*}

\section{Overview}
\label{sec:overview}

Now we introduce an illustrative example to give an overview of our approach~(\autoref{sec:overview-example}).
It is structured into three main phases, \textit{Analysis}~(\autoref{sec:overview-analysis}), \textit{Matching}~(\autoref{sec:overview-matching}), and \textit{Local Reassembly}~(\autoref{sec:overview-reassembly}). The first two phases serve as a prerequisite for the third, which is the centerpiece of the system. Challenges \textbf{C1} and \textbf{C2} are addressed in the initial phases by identifying an over-approximation of the patch and the vulnerable code. 
Challenges \textbf{C3}, \textbf{C4}, and \textbf{C5} are handled in the third phase by 
the local reassembly algorithm utilizing the matching results and control- and data flow analyses from the earlier phases.

\subsection{Running Example}
\label{sec:overview-example}
  \autoref{fig:design_overview} shows the different components of our system. 
  For example, consider a smart camera running a Linux-based firmware that no longer receives (security) updates from the manufacturer. For processing PNG files, the application code in the firmware relies on the open-source library \texttt{libpng 1.0.65}, an outdated and known-vulnerable version of the library.
  Consider a fictional vulnerability (loosely inspired by CVE-2015-8540 in \texttt{libpng}) caused by an improper bounds check of an index variable.
  Specifically, the variable may not be zero or smaller to prevent an out-of-bounds write.
  However, the vulnerable version only checks that the value does not equal zero,
  possibly due to misreading the signedness of the variable.
  Assume that this vulnerability is patched in \texttt{libpng 1.0.66} by fixing the check and additionally assigning a fallback value to it, which was also missing in the vulnerable version.
  In the following, we will explain how our system identifies this patch in binary form and reassembles it in the vulnerable binary.
  
  The input for our patching system consists of two binaries and the name of the affected function. The first binary is the vulnerable library obtained from the firmware binary, \texttt{libpng 1.0.65}, which was compiled by the original manufacturer with unknown settings. The second binary is the patched version of the library, \texttt{libpng 1.0.66}, which we compile ourselves from its publicly released source code. We use this self-compiled \texttt{libpng 1.0.66} to solve the challenge of identifying the parts of the binary that need to be rewritten to patch the vulnerability.
  
  \subsection{Analysis Phase}
  \label{sec:overview-analysis}

  Our approach begins with the analysis phase \circled{1}, which lays the foundation for the following steps. The system recovers the interprocedural control and data flow graphs (CFG and DFG, see~\autoref{sec:preliminaries}) for both binaries. We then use the name(s) of the affected function(s) to identify the corresponding code in the vulnerable binary, by first looking for a matching symbol. If no symbol can be found, we identify the functions using binary code similarity analysis between the patched and vulnerable binaries. 
  \autoref{fig:design_overview} shows a simplified version of the subgraph of the CFG for the affected function in the vulnerable binary of \texttt{libpng 1.0.65} and the corresponding subgraph of the affected function in the patched binary of \texttt{libpng 1.0.66}.

  \subsection{Matching Phase}
  \label{sec:overview-matching}
  
  In the matching phase \circled{2}, we leverage the control- and data flow information gathered in the analysis phase to identify the instructions and data in the patch that need to be reassembled in the vulnerable binary.  
  A binary code similarity analysis determines the basic blocks of the vulnerable function that match the basic blocks of the patched function.
  The set of basic blocks that are not part of a \emph{perfect match} (defined in~\autoref{sec:definitions}) in the patch binary is an over-approximation of the actual patch. Determining this set, which will be written to the vulnerable binary, solves challenge \textbf{C1}. The set of basic blocks not part of a perfect match in the vulnerable binary are then dually an over-approximation of the vulnerability, which will be replaced by the patch, solving challenge \textbf{C2}.
  
  In~\autoref{fig:design_overview}, these basic blocks are visualized by the green nodes in the CFG of the patch binary of \texttt{libpng 1.0.66}. The gray nodes in both CFGs are the matched nodes and the red nodes are unmatched nodes in the vulnerable binary of \texttt{libpng 1.0.65}. The matching reveals that both the vulnerability and the patch consist of two basic blocks each (note that the number of blocks may also differ). If there are no perfectly matched blocks, then all basic blocks of the function will be reassembled. This approach minimizes the amount of code being reassembled while introducing no additional effort compared to reassembling the entire function.
  
  Additionally, the system needs to consider the dependencies between nodes to correctly integrate the patch. Control flow dependencies~(\textbf{C3}) are represented by the CFG; for data flow dependencies (\textbf{C4}), we define the concept of references and matches of references (see~\autoref{sec:definitions}). The last step of this phase is to systematically match such references based on the perfect matches of basic blocks such that there is a one-to-one correspondence between references in the vulnerable and patched versions (see~\autoref{algorithm}).
  
  \subsection{Local Reassembly Phase}
  \label{sec:overview-reassembly}
   
  In the final reassembly phase \circled{3}, the system integrates the identified instructions and data from the patch binary into the vulnerable binary (see~\autoref{sec:rewriting}). 
  The requirements of this phase are inherently tied to the underlying processor architecture, in our case 32-bit ARM, common with IoT devices.
  This architecture features both the ARM instruction set (consisting of 32-bit instructions) and the Thumb instruction set (consisting of mostly 16-bit instructions). Runtime behavior may include switching between those instruction sets.
  Furthermore, PC-relative addressing is essential in both instruction sets, as instructions generally have only limited space for encoding immediate address values.
  The first requirement for successful reassembly is the correct disassembly of machine code as either Thumb or ARM instructions. This is challenging, as the mode of decoding can depend on values computed at runtime~\cite{JiangZLWL020}. While some prior work addresses this problem, perfect disassembly cannot generally be guaranteed~\cite{d-arm}. 
  
  The second requirement is the ability to handle the different instruction sizes possible in Thumb. When not all operands can be encoded in 16 bits during reassembly, the size of a modified instruction may change.
  For instance, in~\autoref{fig:design_overview} we can see a conditional jump (\lstinline|bgt|) to \lstinline|0x4026| in the first green basic block of the patch library at address \lstinline|0x4018|. The relative distance of this jump can be encoded in a 16-bit instruction. However, when this instruction is rewritten to location \lstinline|0xF018| in the vulnerable binary about to be patched, the distance increases and the instruction has to be rewritten to a length of 32 bits. 
  As a result, all following instructions have to be shifted by two bytes, requiring further adjustments to relative addressing. For example, consider the load instruction at address \lstinline|0x401A| in the patch library, which reads data from a PC-relative location.  
  Due to shifts in relative addressing, the offset from the PC value has to be adjusted accordingly, as shown at address \lstinline|0xF01C| in the rewritten library.
  Match \& Mend transparently systematically tracks all required shifts and adjusts offsets as necessary, solving challenge \textbf{C5}. 
  
  After this phase is finished, the system outputs the rewritten vulnerable binary. It now includes the patch and all necessary changes to fix the vulnerability and can be tested for validation.

  \section{Automatic Binary Patching}
  \label{sec:automatic_binary_patching}
  
  In this section, we introduce the core concept and algorithm for local reassembly, solving challenges \textbf{C3} to \textbf{C5}. After defining the usual terminology for program structures~(\autoref{sec:preliminaries}), we introduce our notions of references and matches (\autoref{sec:definitions}), before describing the algorithm itself and how it addresses the challenges~(\autoref{sec:rewriting}).

  \subsection{Preliminaries}
  \label{sec:preliminaries}
  We introduce notation to enable us to define the concept of references and perfect matches of basic blocks.

  \begin{definition}[Basic Block]
  A basic block of a program $P$ is a sequence of consecutive instructions in which flow of control enters at the beginning and leaves at the end, without halt, possibility of branching (conditional jumps), or return, except at the end. A call (\lstinline|bl/blx|) to a function that is expected to return does not end the basic block. We denote the set of all basic blocks of $P$ by $P_{\mathit{BB}}$.
  \end{definition}
  
  Given a program $P$, we denote the set of all \emph{instructions} of $P$ by $P_{\mathit{Ins}}$. A \emph{memory location} of $P$ can be a register, a stack variable, heap variable or a global variable. We denote the set of all memory locations of $P$ by $P_{\mathit{Mem}}$. 
  If $I$ is an instruction of a program $P$, then we denote with $B_I$ the basic block containing $I$.

  \begin{definition}[Definition and use of memory locations]
  Let $v$ be a memory location of $P$. 
  We say that $v$ is \emph{defined} in instruction $I \in P_{\mathit{Ins}}$ if and only if $I$ writes to $v$.
  We say that $v$ is \emph{used} in instruction $I \in P_{\mathit{Ins}}$ if and only if $v$ is read by $I$.
  \end{definition}
  \begin{definition}[Control-Flow Graph] 
  A control flow graph $G_{\mathit{CFG}}=(P_{\mathit{BB}},E_{\mathit{CFG}})$ of a program $P$ is a directed graph with a set of nodes $P_{\mathit{BB}}$ and a set of edges $E= P_{\mathit{BB}} \times P_{\mathit{BB}}$.
  There is a directed edge $(B,C)$ from node $B$ to node $C$, if and only if $C$ can immediately follow $B$ in some execution sequence of $P$:
  
  \begin{enumerate}[(i)]
  \item Jump edge: there is a conditional or unconditional jump from the last statement of $B$ to the first statement of $C$.
  
  \item Call edge: there is a call from an instruction in $B$ to the first instruction of $C$.
  
  \item Fall-through edge: $C$ immediately follows $B$ in the program $P$ and $B$ does not end in an unconditional jump or return.
  \end{enumerate}
  We say that node $B$ is a \emph{predecessor} of $C$ and node $C$ is a \emph{successor} of $B$.
  \end{definition}
  \begin{definition}[Data Flow Graph (DFG)]
  A data flow graph $G_{\mathit{DFG}}=(P_{\mathit{Ins}},E_{\mathit{DFG}})$ of a program $P$ is a directed acyclic graph with a set of nodes $P_{\mathit{Ins}}$ and a set of edges $$E=\lbrace (I,J)_v \mid I,J \in P_{\mathit{Ins}} \text{ and } v \in P_{\mathit{Mem}} \rbrace.$$
  There is directed edge $(I,J)_v$, i.e., a data dependence from $I$ to $J$ with respect to a
  memory location $v$ if and only if there is a non-null path $p$ in the CFG of $P$ from $B_I$
  to $B_J$, with no intervening definition of $v$ and $I$ contains a definition of $v$ and $J$ a use of $v$. 
  \end{definition}

  The local reassembly algorithm in \autoref{sec:rewriting} is designed to be accurate under the assumption of precise CFGs and DFGs. 
  While recovering an exact CFG or DFG of a binary program is a hard and undecidable problem~\cite{theiling00,cfr-vmcai09,acfr-vmcai12}, disassembly of compiler-generated code has become increasingly reliable~\cite{PangYCKPMX21}. 
  We give a detailed explanation of the implementation in~\autoref{sec:implementation} and an empirical evaluation of our prototype in~\autoref{sec:evaluation}.

  \subsection{Locally Scoped Symbolic Mapping}
  \label{sec:definitions}

  Local reassembly of a binary patch implies adding its basic blocks to the CFG of the host binary. To correctly add and integrate these nodes, we must identify all incoming and outgoing edges and their corresponding nodes in the DFG. This process relies on generating a locally scoped symbolic mapping that resolves elements such as PC-relative loads, literal pools, and GOT/PLT entries into concrete addresses or offsets specific to the host binary. To achieve this, we define references and perfect matches of basic blocks.

  \begin{definition}[Reference]
    Let $P$ be a program and $r=(I,J)$, with $I,J \in P_{\mathit{Ins}}$, a pair of instructions of the program. We say that $r$ is a reference if and only if one of the following conditions holds:
    \begin{itemize}
      \item There exists an edge $e=(I,J)_v$, with $I,J\in P_{\mathit{Ins}}$ and $v \in P_{\mathit{Mem}}$, in the data flow graph of $P$, then we call $r$ a data reference.
      \item $I$ is a call or a conditional or unconditional jump instruction and there exists a non-fallthrough edge $e=(B_I,B_J)$, with $B_I,B_J \in P_{BB}$, of the control flow graph of $P$, then we call $r$ a control reference.
    \end{itemize}
     We call $I$ the source  and $J$ the destination of $r$.
  \end{definition}
  
  For minimally invasive patching, we wish to rewrite as few instructions as possible; and consequently, as few basic blocks as possible. 
  To solve this problem, we need to decide whether two basic blocks are semantically equivalent and do not need to be rewritten.
  A fully formal definition of equivalence for basic blocks with respect to the formal semantics of machine code is out of scope for this paper and ultimately unnecessary for our purpose. 
  Instead, we define a pragramatic notion of perfect matches between basic blocks, based on the intuition that two basic blocks should be equivalent if they have the same instructions, allowing for variation in compiler-assigned addresses.

  \begin{definition}[Perfect Match of Basic Blocks]
  Let $P_1, P_2$ be two programs and $B_{P_1}, B_{P_2}$ two basic blocks of $P_1$ and $P_2$, respectively. We say that $B_{P_1}$ and $B_{P_2}$ are a perfect match if and only if     the sequence of instructions in $B_{P_1}$ matches that in $B_{P_2}$; i.e., the sequences have the same length and the i-th instructions have the same opcodes and operands, except for operands used as addresses. 

  \end{definition}

  \begin{definition}[Match of References]
  Let $r=(I,J)$ be a reference in program $P_1$ and $r'=(I',J')$ a reference in program $P_2$. We say that $r$ and $r'$ are a match if and only if at least one of the following conditions holds:
  \begin{itemize}
    \item $B_I$ and $B_{I'}$ are a perfect match of basic blocks.
    \item $B_J$ and $B_{J'}$ are a perfect match of basic blocks. 
  \end{itemize} 
  If $r=(I,J)_v$ is a data reference, then additionally $r'=(I',J')_w$ needs to be a data reference and $v$ and $w$ have to be the same memory location.
  \end{definition}
  These definitions allow us to define our rewriting algorithm (in pseudocode) in the following.

  \begin{algorithm}[t]
    \SetKwIF{If}{ElseIf}{Else}{\textcolor{blue}{if}}{\textcolor{blue}{then}}{\textcolor{blue}{else if}}{\textcolor{blue}{else}}{}
    \SetKwFor{For}{\textcolor{blue}{for}}{\textcolor{blue}{do}}{}

    \SetKwData{Ins}{ins}\SetKwData{InsNew}{insNew}\SetKwData{Ref}{ref}
    \SetKwData{RefDest}{ref.dest}
    \SetKwData{RefVuln}{refVuln}\SetKwData{BS}{bs}\SetKwData{List}{list}
    \SetKwData{Tuples}{tuples}
    \SetKwData{MemLoc}{memLoc}\SetKwData{Value}{value}
    \SetKwData{NewDest}{newDest}
    \SetKwData{Data}{data}
    \SetKwData{Function}{function}
  
    \SetKwFunction{Reassemble}{Reassemble}
    \SetKwFunction{ReplaceRegisters}{ReplaceRegisters}
    \SetKwFunction{GetMatchedRef}{GetMatchedRef}
    \SetKwFunction{BackwardSlice}{BackwardSlice}
    \SetKwFunction{Solve}{Solve}
    \SetKwFunction{WriteData}{WriteData}
    \SetKwFunction{CopyBytes}{CopyBytes}
    \SetKwFunction{LoadDataFrom}{LoadDataFrom}
    \SetKwFunction{AdjustReference}{AdjustReference}
    \SetKwFunction{ReplaceDest}{Replace}
    \SetKwFunction{Save}{Save}
    \SetKwFunction{GetFreeLocation}{GetFreeLocation}
    \SetKwInput{Input}{input}
    \SetKwInput{Output}{output}

    \Input{\Ins(Instructions of \Function)}
    \Output{Rewritten \Function}
    \InsNew $\leftarrow$ \AdjustReference{\Ins}\;
    \eIf{\RefDest not in function}{
      \RefVuln $\leftarrow$ \GetMatchedRef{\Ref}\;
      \If{\Ref is control reference}
        {
          \eIf{\RefVuln is None}
            {
              \NewDest $\leftarrow$ \GetFreeLocation{}\;
              \InsNew $\leftarrow$ \ReplaceDest{\Ins, \NewDest}\;
              \Save{\NewDest, \RefDest}\;
            }
            {
              \InsNew $\leftarrow$ \ReplaceDest{\Ins, \RefVuln}\;
            }
          \Reassemble{\InsNew}\;
        }
        {
          \If{\Ref is data reference}
            {
              \eIf{\RefVuln is None}
                {
                  \Data $\leftarrow$ \LoadDataFrom(\RefDest)\;
                  \WriteData{\NewDest, \Data}\;
                  \BS $\leftarrow$ \BackwardSlice{\Ins, \NewDest}\;
                  \Tuples $\leftarrow$ \Solve{\BS.constraints}\;
                }
                {
                  \BS $\leftarrow$ \BackwardSlice{\Ins, \RefVuln}\;
                  \Tuples $\leftarrow$ \Solve{\BS.constraints}\;
                }
              \For{\texttt{(}\MemLoc, \Value\texttt{$\!\!$)} $\in$ \Tuples}
                {
                   \WriteData{\MemLoc, \Value}\;
                }
              \Reassemble{\InsNew}\;    
            }          
      
        }
    }
    {
      \Reassemble{\InsNew}\;
    }

    \caption{Local reassembly algorithm in pseudocode.}
    \label{algorithm}
  \end{algorithm}
\subsection{Local Reassembly Algorithm}
\label{sec:rewriting}
  
The local reassembly algorithm (Algorithm~\ref{algorithm}) takes the identified basic blocks and references and reassembles them in the vulnerable binary.
The algorithm runs for each affected function and in each function processes the instructions one by one. The shown algorithm therefore only shows the processing for one function, but the algorithm is simply repeated for each affected function in the patch.

Before beginning, we designate a target location for the patch in the vulnerable binary.
For this we create space by extending an executable load segment or, if necessary, by adding a new section \lstinline|.patch| in the vulnerable binary. This section can be extended as needed to accommodate future patches whenever a new one is added.
To redirect the control flow to this new space, we insert a branch instruction at the point where the patch needs to start. 
In the following, reassembling in the vulnerable binary means reassembling at this newly created region in the extended segment.
At this point, we can start the local reassembly algorithm.

\subsubsection*{Shifts}
  The ARM32 Thumb instruction set consists of instructions of 16 and 32 bits length. As described in challenge \textbf{C5}, encoding a longer distance in a jump might lead to an increase of the instruction size. Similarly, this also applies to instructions like the \lstinline{ldr/str} instructions.  This results in a shift in the relative distance between surrounding instructions. Instructions and data are often intermixed due to optimizations for space and execution efficiency and the algorithm needs to ensure that data is not overwritten by the shift in instructions.
  To address this, the algorithm performs two steps, both defined within the function \lstinline{AdjustReference()}.
  Step one is to track the potential candidates that would be affected by a shift. 
  The second step is to adjust references to data inside the patch that might be overwritten, by increasing the relative distance of the reference and effectively moving the data further away such that there is room for shifts. 
  
  For example, instruction \lstinline|ldr r1, [pc,#0x18]| in~\autoref{fig:example} reads data from a memory location that is located inside the patch directly next to instructions. If one of the instructions between the load instruction and the memory location is reassembled from a 16-bit to a 32-bit instruction, the relative distance of the load instruction will be increased, and the data would be overwritten. To prevent this, the algorithm shifts the memory location with its data to a location further away by increasing the value \lstinline|0x18| and tracks this new location.

\subsubsection*{Reference Handling}

\begin{figure}[t]
  \begin{tikzpicture}[font=\sffamily\footnotesize]
  \tikzset{
  hlbox/.pic={
    \node[rounded corners, anchor=north west, 
          fill=blue!40, opacity=0.3, shape=rectangle, 
          minimum width=50mm, minimum height=3.7mm
          ] (-box) at (0.1,-.15) {};
    }
  }
  \node[] (patch_check) at (0,0) {%
  \begin{minipage}{5cm}%
  \begin{lstlisting}[style=assemblyblock]
1000:  push {r4, r5, r6, lr}
1002:  mov  r6, r0
1004:  ldr  r1,[pc, #0x18] 
1006:  mov  r0, r6 
1008:  mov  r5, #0
100C:  add  r1, pc
100E:  bl   png_warning
...          
1020:@\\textit{data}@ 0xF0 0x13 
... 
@\\textit{.png\_warning}@
1288:  push {r4, lr}
128A:  mov  r2, r1
...
2400:@\\textit{ds}@ "zero_length_keyword" 
  \end{lstlisting}%
  \end{minipage}%
  };
  \coordinate (height) at (0,.316); 
  \pic (h1) at ($(patch_check.north west)-2*(height)$) {hlbox};
  \node[xshift=3mm,anchor=west] at (h1-box.east) {\lstinline|r1=0x13F0|};

  \pic (h2) at ($(patch_check.north west)-5*(height)$) {hlbox};
  \node[xshift=3mm,anchor=west] at (h2-box.east) {\lstinline|r1=0x13F0 + 0x1010|};

  \pic (h3) at ($(patch_check.north west)-12*(height)$) {hlbox};
  \node[xshift=3mm,anchor=west] at (h3-box.east) {\lstinline|r2=0x2400|};

  \draw[->,thick] (h1-box.east) 
  to[out=0,in=0]
  node[right]{\lstinline{1020:data}} (h2-box.east);
  
  \draw[->,thick] (h2-box.east) 
  to[out=0,in=0] 
  node[right]{\lstinline{2400:ds}} (h3-box.east);
  
  \end{tikzpicture}    
  
  \caption{Backward data flow slice computed from a memory access instruction. This slice captures the relevant definitions and uses needed to reconstruct equivalent data flow when reassembling the basic block in a different binary.}
  \label{fig:example}
  \end{figure}
  
The algorithm resolves the symbolic mapping of all references of the patch such that the patch is integrated seamlessly into the vulnerable binary.
  We distinguish whether the destination of the reference lies inside or outside the function.
  For references with their destination \emph{inside} the function, no further modifications are necessary, as all addressing is relative. 
  For references with their destination \emph{outside} the function, the algorithm again distinguishes between \emph{control} references and \emph{data} references.
  
  In case of a control reference (\textbf{C3}) the algorithm first invokes the function \lstinline|GetMatchedRef()|, which returns the matched reference, if there is one (e.g., another function in the binary).
  Changing control references to their matched destination connects the patch with the vulnerable binary. The function \lstinline|Replace()| updates the reference's destination to the matched destination address. Then the algorithm reassembles the updated instruction in the vulnerable binary.
  If there is no matched reference, this means that the destination of the reference is not part of the vulnerable binary. By construction, it is thus part of the patch and will be added to the processing queue, allowing the algorithm to reassemble this function as well. 
  Until then, the algorithm modifies the destination of the reference to point to a free location in the vulnerable binary, saving the new address to be used later as a starting point for the affected function.

  For handling of data references (\textbf{C4}), consider the assembly code in~\autoref{fig:example}. Assume that the instructions and data from address \lstinline|1000| to address \lstinline|1040| form the patch that needs to be reassembled in the vulnerable binary. There is a data reference $r=(I,J)_v$ where $I$ is the instruction at address \lstinline|100C| and $J$ is the instruction at address \lstinline|128A|. The memory location $v$ is the global variable \lstinline|2400:ds|. Depending on whether the reference, i.e., the global variable, is matched or not, there are two different paths the algorithm can take.

\subsubsection*{Matched Data References}

  If the reference is matched, the algorithm needs to reassemble the instruction such that the matched memory location (in our example the matched global variable) of the vulnerable binary is used. To do this, function \lstinline|BackwardSlice()| calculates a backward slice starting from the destination of the reference. It traces all instructions that read from the memory location, going backward until it reaches the instruction that writes to it (its definition). The result includes all instructions that use or define the memory location of the reference, as well as all instructions that use or define related memory locations encountered during the slice.
  \autoref{fig:backward_slice} shows the backward slice for the example data reference. In this case it only contains three instructions. 
  The affected memory locations are the registers \lstinline|r1|, \lstinline|r2| and \lstinline|1020:data|
  
  \begin{figure}[t]
  \begin{lstlisting}[style=assemblyblock, backgroundcolor=\color{white}, frame=none]
1004:  ldr  r1,[pc, #0x18]
                      r1  = COPY [0x1020:data]

100C:  add  r1, pc
                      $t1 = INT_ADD 0x100C, 4
                      r1  = INT_ADD r1, $t1

128A:  mov  r2, r1
                      r2  = COPY r1

  \end{lstlisting}
\caption{Instructions and their intermediate representation in the backward slice computed in \autoref{fig:example}.}
  \label{fig:backward_slice}
  \end{figure}

  In a next step, the list of instructions together with the affected memory locations serves as input for the function \lstinline|Solve()|. 
  The function translates the given instructions into an intermediate representation (IR) so that all side effects of the operations are captured. During the translation all used addresses are replaced with the corresponding addresses intended for the reassembled instructions. The function then adds the constraint that the matched memory location is the new destination of the reference.
  
  \autoref{fig:backward_slice} shows the translation of the instructions in the backward slice into IR. Let us assume that the location in the newly created region where the instruction \lstinline[style=plai]|add r1,pc| will be rewritten to is \lstinline|0x130C| and that the matched global variable of \lstinline|0x2400:ds| is located at \lstinline|0x2500|. Then the following equation system is constructed that encodes the relationships among the relevant memory locations:
  \begin{align*}
  r1_a &= \texttt{0x1020:data}\\
  t1 &= \texttt{0x130C} + 4\\
  r1 &= r1_a + t1\\
  r2 &= r1\\
  r2 &= \texttt{0x2500}
  \end{align*}
  Solving this yields \lstinline+r2=0x2500+ and \lstinline+0x1020:data=0x14F0+, which means that the value of the memory location \lstinline{0x1020:data} has to be changed to \lstinline{0x14F0}.
  Note that in Thumb mode the PC register always points to the address four bytes after the current instruction.
  
  The memory locations and the corresponding new values that need to be changed are returned as a list of \lstinline|tuples|. 
  The algorithm reassembles the instruction and writes the new values to the corresponding memory locations with the function \lstinline|WriteData()|. In our example the only memory location that needs a new value is \lstinline{0x1020:data}. 
  
\subsubsection*{Unmatched Data References}

  In case the data reference is not matched, we assume that the value of the memory location of the reference is not part of the vulnerable binary. This may be due to the patch adding a previously unused string to the function, for instance.
  Therefore, the algorithm calls \lstinline|LoadDataFrom()| to load data from the destination of the reference. 
  In the next step, this data is written to a new destination, \lstinline|newDest|, in the extended or added segment. The new destination is positioned at a distance from the instruction to provide sufficient space for rewriting the remaining instructions of the patch.
  Following our example in~\autoref{fig:example}, this means \lstinline|0x2400:ds| has no match and the string \lstinline|"zero_length_keyword"| is loaded from memory location \lstinline|0x2400:ds| and written to a new memory location in the extended segment of the vulnerable binary.
  With this newly created memory location, the algorithm can now proceed as before, calculating a backward slice and using the constraint solver to adapt all memory locations such that address of the newly added data is reached. 
  
  If there is no reference or a reference with destination inside the function, the instruction will be reassembled in the vulnerable binary.
  An example of this can be seen in~\autoref{fig:example}. The instruction at address \lstinline{0x1002} has no reference and will therefore be reassembled to the vulnerable binary.
  
  The algorithm is designed to work correctly under the assumption that the control flow graph and data flow graph are complete and correct. In the following section we will present the implementation of the system and show how we deal with the fact that in general it is not possible to obtain complete and sound control flow and data flow graphs.

\section{Implementation}
\label{sec:implementation}

We now introduce our implementation in the prototype tool Match \& Mend. 
After reviewing the frameworks we employ in our prototype~(\autoref{sec:imp-tools}), we present the implementation following the same phases as in the design, i.e., analysis~(\autoref{sec:imp-analysis}), matching~(\autoref{sec:imp-matching}),
and local reassembly~(\autoref{sec:imp-reassembly}). 

\subsection{Tools and Frameworks}
\label{sec:imp-tools}

For our prototype, we require tools that are able to work with ELF binaries on the ARM32 architecture. 
A number of static binary rewriting tools from the literature would in principle be eligible to implement our approach, and we will discuss these in \autoref{sec:relatedwork}.
In this work, we chose to use the angr~\cite{angr} binary analysis framework to disassemble binaries and to implement our custom analysis and transformations, 
and patcherex~\cite{tool:patcherex} to modify the binary and reassemble instructions. 
Finally, we use lief~\cite{tool:lief} to  modify the structure of ELF files and create sufficient space to add patches to vulnerable binaries.

\subsection{Analysis} 
\label{sec:imp-analysis}

The prototype loads both the vulnerable library binary and the self-compiled patch binary in angr and recovers CFGs for both. Given the size of the binaries and the complexity of constructing an interprocedural CFG for an entire program, we execute the static and fast, but sometimes imprecise, \lstinline|CFGFast| analysis provided by angr for this purpose.  
Because our local reassembly algorithm relies on accurate CFGs for the affected functions, we choose to improve precision locally for those by executing \lstinline|CFGEmulated|.
This computationally more expensive but also more precise analysis recovers the intraprocedural CFG using symbolic execution. 
In the affected functions, we follow up CFG creation by creating the data flow graphs (DFGs) using the \lstinline|DDG| analysis of angr.

After the analysis phase, the system has produced an interprocedural CFG for both binaries, as well as DFGs and intraprocedural CFGs for each affected function. It also maintains a list of references for each affected function, which is used during the matching phase.
Note that, should additional functions be identified as affected and requiring patching during analysis, the analysis phase will be repeated for these, generating precise control- and data flow graphs. 
  
\subsection{Matching} 
\label{sec:imp-matching}
 
To identify the instructions and data that need to be reassembled, we use the BinDiff~\cite{Flake04} implementation of angr to match functions and basic blocks between the vulnerable and the patched version of the open-source library (\emph{Matcher} in \autoref{fig:design_overview}). We are particularly interested in the basic blocks in the affected functions. The resulting matches of basic blocks are then filtered so that only perfect matches of basic blocks (see~\autoref{sec:definitions}) remain.
While BinDiff performs well in most cases, matching can be challenging when binaries have been built with significantly different optimization settings, e.g., when functions have been inlined or loops unrolled~\cite{jia2023inlining}.  
Recent work in the area of binary code similarity analysis promises to improve such shortcomings using machine learning techniques~\cite{asm2vec,duan2020deepbindiff,HaqC21,bin-sim-lessons-learned,jtrans,wang2024clap,dlsp2025-calltargets}. We will investigate the integration of improved matching mechanisms in future work.

At this point, the system has compiled a list of perfect matches and identified basic blocks for the affected function in each binary (\textbf{C1} and \textbf{C2}). They are ordered in ascending address order and the first basic block that is not a perfect match is the starting point for the local reassembly algorithm. Consequently, the last block that is not a perfect match is the end point of the patch. Everything in between forms the patch that needs to be integrated into the vulnerable binary (even if it is a perfectly matched block).

Besides the identification of the patch, the matching of the basic blocks also serves to match references. The prototype compares the references of the vulnerable and the patched version of the open-source library and checks if either their origin or destination is part of a perfectly matched basic block, i.e, they satisfy the definition of a match of references, as described in~\autoref{sec:definitions}. Additionally, we match the global offset table (\textit{.got}), procedure linkage table (\textit{.plt}), and any available exported symbols, if present, to obtain more matched references. Note that in practice, as well as in our experiments, binaries are generally stripped.

The system ends this phase with a list of matched references and a list of basic blocks that need to be reassembled.
  
\subsection{Local Reassembly}
\label{sec:imp-reassembly}

Before beginning reassembly, we reserve space for the patch by extending the first load segment, or, if necessary, adding a new section to the vulnerable binary using lief. Although we could arbitrarily enlarge the segment, it is crucial to avoid disrupting the binary's loading memory layout. For instance, jumps into the \textit{.got} are implemented as relative jumps. Therefore, we must ensure that relative distance of the loading addresses of the \textit{.got} and the code during execution is not altered by our extension of the load segment. Typically, the space created is sufficient to accommodate the patch; however, a check is implemented to add a new section if it is not. This also ensures that subsequent re-patching of the same binary would not run out of available space.
  
Moving forward, the system starts to reassemble the identified basic blocks. The implementation closely follows \autoref{algorithm} to solve \textbf{C3} and \textbf{C4} but has to account for some technical details. Therefore, we will concentrate our exposition on the implementation of the essential functions.
  
The system iterates over the instructions of the patch in ascending address order checking if there is a reference originating from the instruction. 
The function \lstinline|AdjustReference()| adjusts the reference such that there are no conflicts with possible shifts in later instructions. 
It checks whether the instruction reads or writes from a memory location that is part of the patch. If yes, it will reassemble the instruction but increase the distance to the memory location. The new memory location is saved. 
  
  The function \lstinline|BackwardSlice()| implements a worklist algorithm over the DDG to find all statements in the intermediate representation VEX that use a register until its definition. Every register used in these statements is added to the worklist and the process is repeated to find their definitions. The result is a list of VEX statements. 
  
  The function \lstinline|Solve()| processes a list of VEX statements, converting them into an equation system. This process involves iterating through the statements, transforming them into static single assignment form, and integrating them into the equation system. The final constraint added is the new target value. The function solves the equation system and all memory locations are returned with their defining instruction and the new values they require.
  
  \lstinline|LoadDataFrom()| is necessary to load the data from the destination of a reference that does not yet exist in the vulnerable binary. An example for that would be the string \lstinline|"zero_length_keyword"| in~\autoref{fig:example} at address \lstinline+2400+. 
  The function takes as input the address of the string in the patch binary and loads the bytes from this address until the first pair of consecutive null bytes. This heuristic works effectively for strings but may be insufficient for other data types. During our evaluation, we did not encounter the need for more sophisticated methods, but in principle we could leverage known work for type recovery and data structure reconstruction~\cite{LeeAB11,NoonanLC16,typeforge}.
  
  When the local reassembly algorithm is finished, i.e., all instructions have been reassembled, the system must account for shifts and some instructions might enlarge into 32-bit instructions during reassembling (\textbf{C5}). This will happen frequently where code is in Thumb mode, and the system has to encode far jumps. The prototype keeps track of those shifts and for every tracked shift it will add padding bytes after the shifted instruction so that the alignment of the following instructions is correct. Once all instructions are processed, the prototype changes all references affected by the shifts by the number of bytes that were added.

  \section{Evaluation}
  \label{sec:evaluation}

We now evaluate our approach on its ability to effectively patch real-world vulnerabilities in open-source libraries. 
In particular, we answer the following research questions:

\begin{itemize}
\item \textbf{RQ1}: Can Match\&Mend successfully remove vulnerabilities from ELF binaries while preserving existing functionality, given a vulnerability, patch, proof of vulnerability, and test suite?

\item \textbf{RQ2}: Can the system scale to support a wide range of IoT firmware images, rather than being limited to hand-crafted or case-specific scenarios?

\item \textbf{RQ3}: What is the size overhead from patching vulnerabilities?
\end{itemize}

Answering the questions comes with two main challenges. First, we need a vulnerability and corresponding patch, an exploit, and a test suite for the firmware to demonstrate that the vulnerability is properly fixed and that existing functionality remains intact. Second, the evaluation must be scalable to ensure the system is not tailored to just one specific vulnerability or firmware. However, there is a lack of publicly available datasets that meet all these criteria. For this reason, we take a two-fold approach to evaluation.

All experiments were conducted using cross-compilation and emulation on a desktop machine running Ubuntu 22.04.4 LTS on an Intel Core i7-8700
CPU @ 3.20GHz. The patching process took 14 minutes on average; since the system is statically rewriting the binary, this process only needs to happen once.

\subsection{RQ1 Patching Real-World Vulnerabilities}
First, we demonstrate that the system can successfully patch real-world vulnerabilities in open-source libraries, using cases where a proof of vulnerability (PoV) and a corresponding test suite are available. To this end, we use the Magma dataset~\cite{Magma20}, which consists of open-source libraries with known vulnerabilities.

\begin{table}[t]
  \centering
  \caption{Results on Magma.}
  \begin{tabularx}{1.0\columnwidth}{p{21mm}l@{\hspace*{2mm}}p{8mm}rrr}
    \toprule
      \textbf{CVE ID} & \hspace*{-4mm}\textbf{Type} & \textbf{Library} & \textbf{O1}&\textbf{O2} & \textbf{O3} \\ \midrule
      CVE-2018-13785 & \hspace*{-4mm}Integer overflow& LibPNG & \color{green}\ding{51} & \color{green}\ding{51}& \color{green}\ding{51}\\
      (Unspecified) &\hspace*{-4mm}Memory leak & LibPNG & \color{green}\ding{51} & \color{green}\ding{51}& \color{green}\ding{51}\\
      CVE-2015-8781 & \hspace*{-4mm}OOB read & LibTIFF & \color{green}\ding{51} & \color{green}\ding{51}& \color{green}\ding{51}\\
      CVE-2016-9533 & \hspace*{-4mm}OOB read & LibTIFF & \color{green}\ding{51} & \color{green}\ding{51}& \color{green}\ding{51}\\
      CVE-2019-7663 & \hspace*{-4mm}0-pointer dereference& LibTIFF & \color{red}\ding{55} & \color{red}\ding{55} & \color{red}\ding{55} \\
      CVE-2017-11613 & \hspace*{-4mm}Resource exhaustion & LibTIFF & \color{red}{\ding{55}} &  \color{red}\ding{55} &  \color{red}\ding{55} \\
      CVE-2017-0663 & \hspace*{-4mm}Type confusion & LibXML2 & \color{green}\ding{51} & \color{green}\ding{51}& \color{green}\ding{51}\\
      CVE-2017-7375 & \hspace*{-4mm}XML external entity & LibXML2 &\color{red}\ding{55} & \color{green}\ding{51}& \color{green}\ding{51}\\
      CVE-2017-9048 & \hspace*{-4mm}Stack buffer overflow & LibXML2 & \color{green}\ding{51} & \color{green}\ding{51}& \color{green}\ding{51}\\
      CVE-2015-8317 & \hspace*{-4mm}OOB read & LibXML2 & \color{green}\ding{51} & \color{green}\ding{51}& \color{green}\ding{51}\\
      CVE-2016-1762 & \hspace*{-4mm}Heap buffer overread & LibXML2 &\color{red}\ding{55} & \color{green}\ding{51}& \color{green}\ding{51}\\
      CVE-2016-6309 & \hspace*{-4mm}Use-after-free & OpenSSL  &\color{green}\ding{51} & \color{red}\ding{55}& \color{red}\ding{55}\\
      CVE-2017-3735& \hspace*{-4mm}OOB read & OpenSSL  & \color{green}\ding{51} & \color{green}\ding{51}& \color{green}\ding{51}\\
      CVE-2016-6302& \hspace*{-4mm}OOB read & OpenSSL & \color{green}\ding{51} & \color{green}\ding{51}& \color{green}\ding{51}\\

      \bottomrule
  \end{tabularx}
  \label{tab:results}
\end{table}

\subsubsection*{Methodology}
The libraries in the Magma dataset can be compiled with or without vulnerabilities, i.e., for each vulnerability we have (i) a version that represents the vulnerable library obtained from a firmware image, and (ii) a patched version compiled with the provided patch already applied.
Since we want to simulate a real-world scenario, we compile each vulnerability with three different compilation settings, O1, O2, and O3, while the patch is always compiled with O2 and debug information.
The two versions for each compilation setting are given to our prototype. For each vulnerability, the prototype analyzes, matches and locally reassembles the patch in the vulnerable binary, which results in a fixed version of the library. This fixed version is then tested to validate that the patch was successful.
 
We test the fixed binary to check (i) the proof of vulnerability and (ii) correctness. 
For (i), if there is a PoV available, we verify that the vulnerability no longer executes.
Canaries inserted into the vulnerable libraries by Magma indicate whether a given vulnerability is reached and triggered. This allows us to test the patched binary by executing the library with the AFL++~\cite{afl++} fuzzing harness from Magma and the proof of vulnerability as input. We consider the PoV test to be successful if the canary indicates that the vulnerability was not triggered.  
For correctness (ii), we run each fixed binary against its open-source test suite to validate that the patching has not rendered the library unusable. 
We consider the tests successful if the fixed binary passes the same tests as the vulnerable binary does, except for test cases it failed because of the vulnerability. 

We consider a binary to be successfully patched if both the PoV test (i) and the test suite (ii) pass. 
We validated the soundness of our testing methodology by manually inspecting a sample of patched binaries with a disassembler.
Recent work has shown how to validate patches at the binary level using symbolic execution~\cite{WuWWSMB25}, which could help to further automate validation checks.

\subsubsection*{Results}
The Magma dataset contains six open-source libraries and one executable with known vulnerabilities. Since our focus lies on open-source libraries for ARM-based IoT-devices, we only consider the libraries in the dataset that we could successfully compile for this architecture with the correct configurations, i.e., LibPNG, LibTIFF, LibXML2, OpenSSL, and SQLite. 
Since there is no publicly available binary-level test suite for SQLite, and we did not receive a reply from developers upon request for an academic license, we excluded SQLite.
Out of those libraries we selected all 14 vulnerabilities that are limited to one function and where MAGMA provided a POV. 
The resulting set covers a broad variety of vulnerability types, as shown in~\autoref{tab:results}. 
All libraries were compiled using the gcc cross-compilation toolchain \lstinline|arm-linux-gnueabihf-gcc|.

The adapted MAGMA dataset contains 14 libraries with different vulnerabilities, each compiled in three versions, resulting in 42 vulnerable libraries in total. Our prototype was able to process all of these. The detailed results are shown in~\autoref{tab:results}. 
Out of the 42 libraries, 32 (76\%) have been patched successfully; for the remaining 10, patching was unsuccessful. Two of these were not patched correctly because their optimization level (O1) led to wrong matches. 
Similarly, matching failed on levels O2 and O3 in one vulnerability in OpenSSL.
The other six cases involve two vulnerabilities in LibTIFF. For each vulnerability, patching failed across all optimization levels, because the backward slicing was unable to correctly follow data flow through the stack in the patched version.

The limitations in function matching indicate that replacing the BinDiff implementation with more accurate machine-learning based systems could further improve the end-to-end performance of our system.

Despite limitations from BinDiff and fundamental challenges in static analysis of binaries, Match\&Mend  successfully patches the overwhelming majority of cases, demonstrating that it could improve security in practice.

\subsection{RQ2 Patching IoT Firmware Images}
 
The second part of the evaluation focuses on testing Match\&Mend in a realistic deployment setting. Using firmware images from the Karonte dataset~\cite{RediniM0SCSKV20}, we demonstrate that the prototype can operate effectively in real-world conditions. 
The dataset does not provide information about known vulnerabilities, let alone proofs of vulnerabilities, so we rely on public CVE databases for identifying candidate vulnerabilities and test suites for validating functionality.
In combination with the evaluation of RQ1, this demonstrates that our approach is effective across a broad range of real-world firmware scenarios.

\subsubsection*{Methodology}
The Karonte firmware dataset consists of 49 Linux-based firmware images from four major IoT vendors that make the firmware of their devices available for download: NETGEAR, TP-Link, D-Link, and Tenda. Since the dataset neither provides information about included open-source libraries nor known n-day vulnerabilities, we use the CVE Binary Tool~\cite{tool:CVE-bin-tool} to identify open-source libraries and their versions in the firmware images. 
The CVE Binary Tool is a Python-based open source scanning tool maintained by Intel, which tries to automatically identify libraries and other types of software, as well as their specific version, and then matches this information with published CVE information.
The identification mainly relies on relevant product and version strings in the scanned file.
We excluded all files from the list of supposedly vulnerable files that were not compiled for a 32-bit ARM architecture.
Furthermore, we selected six CVEs  based on their relevance, i.e., the number of affected firmware images in the Karonte dataset, and on the ease of cross-compiling the corresponding libraries (libpcap, zlib, libpng, libflac) in all affected versions.

We built a script to prepare patched versions of the selected test subjects. 
Where possible, we automatically identified the next unaffected version for a given CVE, fetched the source code, and compiled it using a toolchain built with \texttt{buildroot}~\cite{tool:buildroot}.
The toolchain uses either glibc~\cite{glibc} or uClibc-ng~\cite{uclibc-ng}, depending on the loader in the firmware of the affected binary. 
In the case of CVE-2019-15165, the delta of the version fixing the CVE was too great, requiring manual backporting of the patch as if creating a hotfix for a previous source code release.
Note that manual backporting hot patches at the source level is common practice in security-critical projects; because of its effectiveness for improving security, several research projects attempt to expand the scope of backporting by largely automating the process~\cite{ShariffdeenGDTL21,Yang0XSJZ23}.

Additionally, our script fetches and compiles the source code of the vulnerable library (i.e., the same version as in the firmware) to
provide an environment to run the unit tests for the version that Match \& Mend fixed.
As we build the patched versions of binaries ourselves, we are free to choose compilation options. In particular, we can also build multiple versions with different compilation options to improve the quality of matching. For our experiments, we use the optimization levels Os (for size, common with embedded systems) and O2; for libpng, O2 yields best results, whereas Os is sufficient with all other targets. The process could be further improved using configuration inference, as demonstrated by OSSPatcher~\cite{DuanBJAXISL19}. However, for our evaluation, we relied on lightweight matching, which was effective in our setting.

With these modifications, the CVE Binary Tool is able to supply the patching prototype with most of the information it needs to  patch a vulnerable binary, i.e., the vulnerable binary itself, a compiled patched version, and a compiled affected version, as well as its source directory. The only missing information is the affected function, which we deduce from the CVE description and the source code of the affected version. Since CVE descriptions can be unclear or may lack important information~\cite{KuehnBW021}, this is the most time-consuming part of the process. 
For example, a statement like "all versions before \textit{1.30.9}" can be interpreted in multiple ways, such as referring to versions from \textit{1.30.0} to \textit{1.30.9}, or from \textit{1.0.0} to \textit{1.30.9}. Moreover, CVEs tend to describe the impact of a vulnerability rather than its root cause. Since Match\&Mend is designed to patch n-day vulnerabilities, we generally rely on the assumption that the affected function is known. While CVE entries are the most obvious source of this information, bug reports or commit messags are equally valid sources. 
Fortunately, the importance of vulnerability databases has become increasingly clear to vendors and policy makers, and we expect the quality of such reports to improve. 
Overall, we were able to identify the affected function for six CVEs in different versions of different libraries, which we then patched and tested.

To test a patched library, we cross-compile the test suite for the affected library, if available. If no test suite exists or if it cannot be cross-compiled, we write manual test cases for the affected functions. All tests are executed within the filesystem of the original firmware image.
Although this does not allow us to directly confirm that a vulnerability has been fixed, we know after testing that a patch was applied and that the patched library remains functional. 

\begin{table}[t]
  \centering
  \caption{Results on real world firmware vulnerabilities. ``Affected'' are distinct binaries in different firmware images.}
  \begin{tabularx}{\columnwidth}{Xlrrrr}
  \toprule
  \textbf{CVE ID} &  \textbf{Library} & \hspace*{-2mm}\textbf{Affected} & \hspace*{-2mm}\textbf{Patched} & \hspace*{-2mm}\textbf{Passed} & \hspace*{-2mm}\textbf{Success} \\
  \midrule
  CVE-2019-15165\hspace*{-2mm}  &  libpcap   & 7 & 3 & 3 & 43\%  \\
  CVE-2016-9840   &  zlib      & 33 & 33  & 33 & 100\% \\
  CVE-2016-9842   &  zlib      & 24 & 24 & 24 & 100\%\\
  CVE-2016-9843   &  zlib      & 33 & 33 & 33 & 100\%\\
  CVE-2016-10087\hspace*{-2mm}  &  libpng    & 7 & 6 & 6 & 85\%\\
  CVE-2020-22219\hspace*{-2mm}  &  libflac & 27 & 27 & 27  & 100\%\\
  \midrule
  Total &  & 131 & 126 & 126 & 96\%\\
  \bottomrule
  \end{tabularx}
  \label{tab:realworld}
\end{table}

\subsubsection*{Results}  
Our system can successfully patch 126 out of~131 vulnerabilities across all 30 ARM32 firmware images in the Karonte dataset. As shown in~\autoref{tab:realworld}, whenever the system was able to apply a patch, the corresponding library also passed its test suite. In the five remaining cases, Match\&Mend was unable to patch the vulnerability because the underlying analysis tool, angr, failed to construct an interprocedural control flow graph for the affected library.
  These results indicate that our approach is effective in practice even when dealing with libraries that are stripped of section headers and symbols and compiled with unknown configurations. Overall, Match\&Mend was able to patch all ARM32 firmware images in the Karonte dataset.
  
\subsection{RQ3 Size Overhead}

Another important aspect of the performance of the system is its ability to adhere to the micro-patching strategy, i.e., to patch only the affected functions and not the entire library.
The design of Match\&Mend is based on basic block matching, enabling the system to patch individual basic blocks to remove vulnerabilities. In practice, however, the system often needs to patch more than a few basic blocks either due to matching difficulties or because the patch itself spans multiple blocks. Often, the system would end up patching entire functions. 

\begin{figure}
  \centering
  \includegraphics{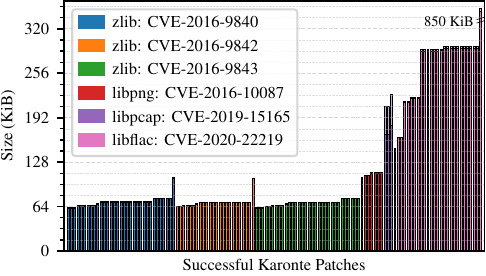}
  \caption{Histogram showing the size of each vulnerable library and the additional size of its patch.}
  \label{fig:file_size_changes}
\end{figure}

Nevertheless, the analysis is identical whether we use basic blocks or functions as the fundamental units for matching. 
Using basic blocks provides us with the opportunity to patch very small units whenever possible, but can be extended seamlessly for larger areas of the binary. 
\autoref{fig:file_size_changes} shows the sizes of the vulnerable binaries and the bytes needed to patch them. 
The observed sizes of the identified patches support the assumption that security patches tend to be small, supporting the micro-patching strategy ranging from one to 34 lines of code. While the current prototype does not yet optimize the use of available space and always adds the patch segment at a fixed size, the average patch sizes observed in real-world firmware images range from 
5 to 46 KiB or 9 to 66 basic blocks, compared to library sizes between 65 and 850 KiB. The larger patches correspond to cases where larger amounts of data had to be copied; the only outlier is a library that includes debug information, which is uncommon with production firmware.

As the patches are added to a new section of the binary, if required, we can in principle patch an arbitrary number of vulnerabilities over time. While theoretically the file size could grow larger than available memory, the patch sizes shown in~\autoref{fig:file_size_changes} indicate that this is unlikely to occur within the lifespan of a device.

Thus, we can conclude that Match\&Mend achieves the goal of micro-patching, by only modifying a very small fraction of the code in the affected libraries. This presents a significant improvement over replacing entire libraries with fixed versions, which is error-prone due to the wide variety of compiler options, build configurations, and potential custom adaptations.

  \section{Limitations}
  \label{sec:limitations}
  
In the following, we discuss some limitations of the implementation and evaluation of our system.

The accuracy of our reassembling process depends on the quality of the generated control flow graph (CFG) and data flow graph (DFG). In general, the current implementation strikes a reasonable balance between precision and performance, but it has some limitations. For instance, data flows that pass through the stack are not yet tracked reliably. Since our use case allows partial access to source-level information and limiting the scope locally, this context could be leveraged to improve the precision of CFG and DFG generation in targeted areas.

Identifying the affected functions in the source code is a prerequisite for our approach; however, locating the corresponding functions in the vulnerable binary presents a separate challenge. As this is not the main focus of our prototype, we opted for a simple and practical solution. Searching for symbol names is a reasonable first step, but it fails when the binary is stripped, a common scenario, particularly in embedded systems. In such cases, we fall back on a BinDiff based implementation using angr, which performs well in many cases. Nevertheless, matching functions becomes difficult when the target has been inlined or compiled with significantly different settings. While our current method works in most cases, it is limited in handling these edge scenarios. The field of binary code similarity analysis offers many improvements~\cite{duan2020deepbindiff}, and integrating machine learning models tailored to this task could enhance locating the function.

Local Reassembly, as implemented in Match\&Mend, integrates code in binary form into an existing binary. To match substantial parts of the patch binary to the vulnerable binary, it is beneficial to approximate the original compilation settings as closely as possible. This observation is also supported by our MAGMA evaluation, where several failures were caused by incorrect function matches across different optimization levels. We do not require a fully exact reconstruction of the build environment, which would only be necessary when rebuilding an entire library, but recovering key compiler parameters has a strong impact on matching quality.
One particularly influential parameter is the compiler optimization level. Our current approach is pragmatic: we compile the patched source using several candidate optimization levels, typically four or five, and then select the configuration that yields the highest similarity to the original binary. This procedure mitigates limitations in our function matching, because a binary compiled with an appropriate optimization level more closely resembles the original binary

\begin{table*}[!h]
  \centering
  \caption{Comparison of different binary rewriting and patching tools.}
  \begin{tabularx}{\textwidth}{XXlXXllccccc}
  \toprule
  \textbf{Name} & \textbf{Rewriting} & \textbf{Architectures} & \textbf{Modification} &  \textbf{Main Goal} &\textbf{C1} &\textbf{C2} &\textbf{C3} &\textbf{C4} & \textbf{C5} \\ \midrule
  RetroWrite~\cite{DineshBXP20} & Static & x86\_64 & Global & Instrumentation & \circlet & \circlet & \circlet & \circlet & \circletfill \\
  Repica~\cite{HaJO18}   & Static & ARM32, AArch64 & Global & Instrumentation & \circlet & \circlet & \circlet & \circlet & \circletfill \\
  Egalito~\cite{egalito}& Static & x86\_64, AArch64 & Global & Instrumentation & \circlet & \circlet & \circlet & \circlet & \circletfill \\
  DDisasm~\cite{Flores-MontoyaS20}  & Static & x86\_64 & Global & Disassembly & \circlet & \circlet & \circlet & \circlet & \circletfill \\
  ICFGP~\cite{MengL21}  & Static & x86\_64, ppc64le, AArch64 & Local & Instrumentation & \circlet & \circlet & \circlet & \circlet & \circlet \\
  E9Patch~\cite{DuckGR20}  & Static & x86\_64 & Local & Instrumentation & \circlet & \circlet & \circlet & \circlet & \circlet \\
  MultiVerse~\cite{BaumanLH18} & Static & x86\_32 & Global & Instrumentation & \circlet & \circlet & \circlet & \circlet & \circletfill \\
  Armore~\cite{BartolomeoMP23}    & Static & AArch64 & Global & Instrumentation & \circlet & \circlet & \circlet & \circlet & \circletfill \\
  RevARM~\cite{KimKCKSZX17} & Static & ARM32 & Local & Instrumentation & \circlet & \circlet & \circlet & \circlet & \circletfill \\
  BinPatch~\cite{HuZG19} & Static & x86\_64 & Local & Patching & \circletfillhr & \circletfillhr & \circletfillhr & \circletfillhr & \circlet \\
  OSSpatcher~\cite{DuanBJAXISL19} & Dynamic & ARM32, AArch64, x86\_64 & Local & Patching & \circletfill & \circletfill & \makebox[6pt][l]{\circletfill$^{\textbf{\textit{ d}}}$} & \makebox[6pt][l]{\circletfill$^{\textbf{\textit{ d}}}$} & \circlet \\
  \textbf{Match \& Mend}& Static & ARM32 & Local & Patching & \circletfill & \circletfill & \circletfill & \circletfill & \circletfill\\
  \bottomrule

  \end{tabularx}
  \label{tab:relatedwork}
\end{table*}

\section{Related Work}
\label{sec:relatedwork}
Now we discuss existing work for static binary rewriting and binary patching, as shown in~\autoref{tab:relatedwork}, and whether they offer a solution for the identified challenges \textbf{C1} to \textbf{C5}. 
To the best of our knowledge, Match \& Mend is the only tool that solves all challenges for Linux-based ARM32 binaries.

  Automated binary patching of vulnerabilities requires precise modifications to address specific security issues, often replacing or fixing faulty logic. Existing techniques for binary rewriting merely shift the problem of patching to a different level by creating modifiable intermediate representations or reassembleable assembly while not providing a technique for systematically integrating patches into the binary.
  Most work is focused on rewriting binaries for analysis or monitoring purposes. While the core challenges of binary rewriting remain similar independent of the rewriting goal, techniques for instrumentation are not necessarily suitable for patching vulnerabilities. The key difference lies in the type of binary code added by the binary rewriting process. Instrumentation, for example, refers to adding code for dynamic analysis, usually in a way that does not alter the intended functionality of the original program.
  Still, the core challenges of binary analysis like control flow recovery, data flow analysis or symbolization remain and do not change with the goal of the binary rewriting. 
  \autoref{tab:relatedwork} provides an overview of binary rewriting and patching tools, their focus and target architecture and whether they address the challenges with a full solution (\circletfill), a partial solution (\circletfillhr), or a solution specific to the dynamic setting (\circletfill$^\textbf{\textit{ d}}$). In the following we will discuss the work in more detail.
  
  Overall, we can classify all approaches into two categories based on their impact on the binary. The first category is global rewriting, where the entire binary is transformed. The second category is local rewriting, which focuses on modifying only specific parts of the binary. Match\&Mend follows a local micro-patching strategy.
 
  \subsection{Global Rewriters}

  Tools in the first category are commonly designed to instrument binaries for purposes such as fuzzing, hardening, or analysis. While they modify the entire binary and are not specifically tailored for patching vulnerabilities, their binary rewriting capabilities could, in principle, be adapted to our use case. However, their global modifications conflict with our minimally invasive and localized approach. In the following, we analyze how these tools address challenges \textbf{C1} to \textbf{C5}.
  
  \textbf{RetroWrite}~\cite{DineshBXP20} and \textbf{Armore}~\cite{BartolomeoMP23} create reassembleable assembly through symbolization for 64-bit x86 and AArch64 PIE binaries, respectively. Leveraging relocation information of position-independent binaries, they do not rely on heuristics to symbolize the binary. Since their focus lies more on instrumentation and not on patching known vulnerabilities, they do not offer solutions for patch or vulnerability identification (\textbf{C1} and \textbf{C2}). While the authors show that Armore can be used to patch vulnerabilities, doing so requires an expert who manually patches the reassembleable assembly. 
  Hence, there is no approach to integrate control or data flow of a patch into the binary. Encoding modified instructions is solved implicitly by RetroWrite and specifically Armore  by symbolizing the entire binary. Both tools work on Linux-based systems but are not compatible with ARM32 binaries.
  
  \textbf{Repica}~\cite{HaJO18} is a static binary instrumentation technique that symbolizes the entire binary. Using a value-set-analysis specialized for position independent ARM binaries, Repica identifies and processes relative addresses.
  Any changes to be made need to be specified manually in an instrumentation specification, shifting the problems of locally integrating a patch to another level and hence not addressing challenges \textbf{C1} to \textbf{C4}. Challenge \textbf{C5}, encoding modified instruction is also implicitly solved by symbolizing the entire binary. The tool is compatible with ARM32 and AArch64 binaries.
  
  \textbf{DDisasm}~\cite{Flores-MontoyaS20} is a static binary rewriter, developed in Datalog for 32-bit x86 binaries, and since publication, also ARM32. It statically recovers control- and data flow information and uses reassembly heuristics optimized through the use of datalog. The entire binary is symbolized focusing on analysis and monitoring purposes rather then automated patching. Challenges \textbf{C1} to \textbf{C4} are not addressed by DDisasm but symbolizing the entire binary implicitly solves challenge \textbf{C5}.
  
  \textbf{Egalito}~\cite{egalito} recompiles 64-bit binaries (x86 or ARM) for rewriting by first lifting them to an intermediate representation. This global approach shifts the challenges \textbf{C1} to \textbf{C4} to the level of the intermediate presentation but does not solve them. Similar to the other global approaches the encoding of modified instructions is implicitly solved by the recompilation process. 
  
  \textbf{MultiVerse}~\cite{BaumanLH18} statically rewrites x86 binaries without heuristics. It completely disassembles instruction sequences starting from every byte offset in the binary's text section constructing a set of instructions that contains a subset with all valid instructions. Indirect control flow is rewritten using a table mapping new addresses to the original ones. MultiVerse is then able to relocate all instructions to any other location without changing the original control flow. The main purpose of their technique is disassembly and instrumentation and does not offer a solution for \textbf{C1} to \textbf{C4}. For x86 binaries they solve challenge \textbf{C5}, but this approach is not applicable for ARM architectures.

  \subsection{Local Rewriters}

  The second category of tools focuses on local rewriting and includes our own tool, Match \& Mend.
  
  \textbf{ICFGP}~\cite{MengL21} rewrites 64-bit binaries (x86\_64, AArch64, ppc64le) locally by inserting trampolines.  The tool is implemented on top of Dyninst~\cite{BernatM11} and implements an analysis to find the best places to put trampolines. It uses the trampolines to jump to a specially crafted instrumentation area and then incrementally tries to reduce the amount of trampolines while changing all necessary control flow so that the original semantics of the binary can still be executed. The purpose of their tool is instrumentation. In their evaluation they use empty instrumentation to show that their tool does not break the binary. Challenge \textbf{C1} to \textbf{C5} are not addressed.

  \textbf{E9Patch}~\cite{DuckGR20} follows a minimally invasive local approach by inserting trampolines into x86\_64 binaries so that binary code can be injected into it. It works without recovering control flow and on a very low level. As input, it takes an unpatched binary, disassembly information, a set of patch instruction locations and trampoline templates. This input needs to be provided by a user and does not offer an automated solution for challenges \textbf{C1} to \textbf{C5}.

  \textbf{RevARM}~\cite{KimKCKSZX17} instruments ARM (both ARM32 and AArch64) binaries without limitation on the target platform. It takes an input binary and instrumentation specifications such as security policies to enforce and target embedding/extraction locations and generates an instrumented binary. The instrumentation specification requires information about the binary, the vulnerability and a possible binary code fix that is not easily available for IoT-firmware. Our system tries to automate the process of creating this information and therefore does not require a manually created instrumentation specification. Thus challenges \textbf{C1} to \textbf{C4} are not addressed, but it solves challenge \textbf{C5} by inserting NOP statements to keep alignment and adjusting the encoding of references when necessary.
   
  \textbf{BinPatch}~\cite{HuZG19} is an automated binary patching tool for IA32 binaries. Similar to our system the tool uses binary code similarity analysis together with a description of the vulnerability to identify patch and vulnerability (challenges \textbf{C1} and \textbf{C2}). Additionally, BinPatch requires a known exploit as a trigger for the vulnerability.
  It uses this exploit as an input case for symbolic execution. BinPatch executes a self-compiled version of the patch and symbolically executes the vulnerable firmware version with this input. Afterwards, it compares the results of the two executions and uses the gained information to rewrite the vulnerable function. This approach limits the applicability of BinPatch to vulnerabilities that have a known exploit. As we have seen, even for a curated dataset like Magma this is the case for only about 45\% of the vulnerabilities.
  Furthermore, BinPatch is not able to patch vulnerabilities that have the same CFG as their patch and is limited to vulnerabilities that are patched by adding new conditions or modifying the original ones. It can only perform the patching if the data structure of the vulnerable and the patch function is the same.
  Under those constraints on control flow and data flow, BinPatch solves challenges \textbf{C3} and \textbf{C4}. Challenge \textbf{C5} is also solved for the targeted IA32 architecture.
  
  \textbf{OSSPatcher}~\cite{DuanBJAXISL19} is the most closely related work to ours, as shown in \autoref{tab:relatedwork}, and is capable of dynamically patching known vulnerabilities in binaries on Android. 
  Like Match\&Mend, OSSPatcher leverages information from publicly available source code to guide its binary patching process.
  It identifies both the patch and the vulnerable code using CVE information and function-level matching, thereby addressing challenges \textbf{C1} (on the source level) and \textbf{C2}. Once a source-level patch is identified, OSSPatcher performs a feasibility analysis to determine whether the patch can be applied.   
  To generate a patch, OSSPatcher compiles a stub library and a patch library, with dependencies to both the original vulnerable libraries and the stub. Instead of employing static binary rewriting to solve challenges \textbf{C3} (Control Flow Integration) and \textbf{C4} (Data Flow Integration) this design delegates the resolution of references to the linker and loader at runtime. 
  To deploy patches, OSSPatcher uses \texttt{ptrace} to monitor Android's Zygote process and injects the libraries in memory on matching \texttt{ope}n and \texttt{mmap} system calls.  
  While this is an elegant design within the lifecycle of Android applications, it is unsuitable for our use case. A similar deployment mechanism for general embedded Linux distributions would require (binary) patches to the kernel, which goes against our premise of minimizing possibly damaging side effects. 
  Furthermore, the creation of separate libraries causes significant space increases per patch, with the evaluation reporting up to 80 KB. While OSSPatcher is not publicly available for direct comparison, this would, in our dataset, amount to a size overhead of up to 125\% per patch for the smallest library.

  The authors of OSSPatcher acknowledge that static binary rewriting could be an alternative implementation path. However, pursuing this would necessitate addressing challenges \textbf{C3} (Control Flow Integration) and \textbf{C4} (Data Flow Integration) differently, as such information can only be resolved dynamically by the loader. For static binary rewriting, OSSPatcher does not provide mechanisms to solve these challenges, underscoring the need for a dedicated solution such as Match\&Mend, which is explicitly designed to address them in the static binary rewriting context.

\section{Conclusion}
\label{sec:conclusion}
In this work, we were able to show that minimally invasive local reassembly can be a viable solution for patching (n-day) vulnerabilities in accessible IoT firmware, without vendor support. To this end, we implemented Match \& Mend, a prototype that automatically reassembles patches for known vulnerabilities in ARM 32-bit firmware binaries. The system is based on a novel three-phase approach that includes binary analysis, matching and local reassembly. It implements a micro-patching strategy that utilizes a locally scoped symbolic mapping.

The evaluation of our approach on known vulnerabilities and real world firmware images highlights the practicality and potential of binary-level patching. By focusing on minimal invasiveness and reducing the risk of breaking changes, our system achieves a high success rate at minimal size overhead.
Moreover, our approach will benefit from further advances in binary code similarity detection, automated backporting, and patch verification, which are all active areas of research; therefore, we consider minimally invasive local reassembly a promising approach towards enhancing the security and extending the lifespan of IoT devices.

\section*{Acknowledgment}

This work was supported by the Bavarian State Ministry of Science and Arts (BayStMWK), under project ``ForDaySec: Security in Everyday Use of Digital Technologies (fordaysec.de)'' of the Bavarian Research Association.

\IEEEtriggeratref{54}
\bibliographystyle{IEEEtran}
\bibliography{IEEEabrv,bibliography.bib}

\end{document}